
\input harvmac.tex
\input epsf

\def\figin{\epsfcheck\figin}\def\figins{\epsfcheck\figins}
\def\epsfcheck{\ifx\epsfbox\UnDeFiNeD
\message{(NO epsf.tex, FIGURES WILL BE IGNORED)}
\gdef\figin##1{\vskip2in}\gdef\figins##1{\hskip.5in}
instead
\else\message{(FIGURES WILL BE INCLUDED)}%
\gdef\figin##1{##1}\gdef\figins##1{##1}\fi}
\def\DefWarn#1{}
\def\figinsert{\goodbreak\midinsert}
\def\ifig#1#2#3{\DefWarn#1\xdef#1{fig.~\the\figno}
\writedef{#1\leftbracket fig.\noexpand~\the\figno}%
\figinsert\figin{\centerline{#3}}\medskip\centerline{\vbox{\baselineskip12pt
\advance\hsize by -1truein\noindent\footnotefont{\bf Fig.~\the\figno:}
#2}}
\bigskip\endinsert\global\advance\figno by1}

\lref\RLL {E. Lifshits, L. Pitaevski, {\sl Physical Kinetics},
Pergamon Press, (1982)}
\lref\RDrei {H. Dreicer, {\sl Phys. Rev.} (2) 115 (1959) }
\lref\RGur {A.V. Gurevich, {\sl Sov. Phys. JETP} {\bf 11} (5) (1960) 1150 }
\lref\RKogan {I. Kogan,  private communications}
\lref\RLa {V. Berestetskii,  E. Lifshits, L. Pitaevski,
 {\sl Quantum Electrodynamics}, Pergamon
Press, (1981)}
\lref\Rplasma {L. McLerran, {\sl Rev. Mod. Phys.} {\bf 58} (4) (1986) 1021}

\font\title = cmr10 scaled \magstep 2
\font\subtitle = cmr10 scaled \magstep 1

{ \centerline{}
{ \bf October 1994 \hfill PUPT-1506 \break}
{ \null \hfill hep-ph/9411231}
\vskip 30 pt
\centerline { \title  Runaway Quarks}
\baselineskip = 50 pt
\centerline {\subtitle V. Gurarie\footnote{${}^{\dagger}$}
{gurarie@puhep1.princeton.edu}}
}
\vskip 20 pt
\centerline {\subtitle \sl Department of Physics}
\centerline {\subtitle \sl Princeton University}
\centerline {\subtitle \sl Princeton, NJ 08544}
{\baselineskip=150 pt
\centerline {\subtitle \bf Abstract }}
\vskip 10pt

{ When heavy nuclei collide, a quark-gluon plasma is formed.
The plasma is subject to strong electric field due to the
charge of the colliding nuclei. The electric field can influence
the behavior of the quark-gluon plasma. In particular, we might observe
an increased number of quarks moving in the direction of that field,
as we do in the standard electron-ion plasma.
In this paper we show that this phenomenon, called
the runaway quarks, does not exist.
}

\vfill
\eject

The phenomenon of the runaway electrons in the electron-ion plasma has been
discovered by Dreicer \RDrei \ in 1959 and was later described  by
Gurevich \RGur\ in 1960.
They were interested in what happens to the distribution
function of the electrons in plasma if there is a weak external homogeneous
electric field applied. The somewhat unexpected discovery was that regardless
of how weak this field was, the distribution function of the fast electrons
was found to be strongly distorted.

The essence of the phenomenon lies in the fact that the fast electrons
travel  large distances on the average between collisions. While the
slow electrons will give away all the momentum they acquired due to the
motion in the direction of the electric field at the instant of
collision, the fast electrons acquire a lot of energy which can not
be given away by collisions. The faster they travel, the less
important the collisions become for them. As a result, we observe the
flux of fast electrons flowing in the direction of the electric field.
The bulk of the electrons which are still distributed according to
Maxwell distribution play the role of the ``heat bath'' with the
spring of the electrons in the direction of bigger energies in the
momentum space.

Recently it has been remarked by I.Kogan \RKogan \  that somewhat
similar physical setting is achieved in a completely different
situation, namely at the collision site  of
heavy nuclei. After their collision, a quark-gluon plasma is believed to be
formed ( see ref. \Rplasma \ and references therein)
with the temperature around 200 MeV. The plasma consists
of mainly light quarks and antiquarks which may be considered to be
ultrarelativistic, that is massless, particles under such a temperature.
On the other hand, the charge of the original nuclei can not be
destroyed in the process, but rather it surrounds the plasma
putting it into a strong electric field. There is a possibility of
light quarks running away in the very same way as the electrons in
plasma.

In this paper we consider this phenomenon in detail and show that due
to the effects of relativistic mechanics the quarks will not run away
and the only influence the electric field will have on them will be
a slight distortion of the distribution function.

To do it, we need to derive the kinetic equation of the quarks in
quark-gluon plasma. The main interaction mechanism here is the
exchange of a single gluon between quarks since at such  high temperatures
the perturbative QCD becomes applicable. The resulting interaction is
very similar to the standard QED photon exchange, so the force between
quarks will be of the same functional form as the electromagnetic force between
electrons in plasma. This brings us to the conclusion that we should use
the Landau collision integral (see \RLL) for the kinetic equation.

In deriving the kinetic equation of the electrons in plasma, Landau noticed
that the collisions with a small momentum transfer become important because
of the slow decrease of the Coulomb force with distance. It allows to
consider the collision process as a diffusion in the momentum space.

Let us review the derivation of Landau Collision Integral following \RLL \
but not assuming that the plasma we are considering is nonrelativistic
having in mind the application to quark-gluon plasma.

The collision integral can be expressed in terms of the divergence of
the flux of the particles of plasma

\eqn\Ebasic { {d f \over d t}= - { \partial s_\alpha \over
\partial p_\alpha} }
where $f$ is the distribution function, $p_\alpha$ is the momentum
 and $s_\alpha$ is the flux. Let
us find that flux in terms of the distribution function.

Let
$$ w f( {\bf p}) f^\prime ({\bf p^\prime})d^3q d^3p^\prime $$
be the number of collisions per unit time
between the particles of the momentum ${\bf p}$
and ${\bf p^\prime} $, {\bf q} being the momentum transfer. We remember
that we will be interested in collisions of high energy runaway particles
with the thermal particles, so we denoted the distribution of high
energy particles by $f$ while denoting the Maxwell distribution by $f^\prime$.

We will write down
$w$ as a function of ${\bf p}$, ${\bf p^\prime}$, and ${\bf q}$ in the form

$$ w \left( {\bf p}+ {{\bf q} \over 2}, {\bf p^\prime} -{{\bf q} \over 2} ,
 {\bf q} \right). $$
Here the first argument is half sum of the initial and final momenta of the
first particle, while the second one is half sum of the initial and final
momenta of the particle the first one collides with.

The evident property of $w$ is

\eqn\Eprop {w \left( {\bf p}+ {{\bf q} \over 2}, {\bf p^\prime} -{{\bf q} \over
2} ,
 {\bf q} \right) =
w \left( {\bf p}+ {{\bf q} \over 2}, {\bf p^\prime} -{{\bf q} \over 2} ,
- {\bf q} \right)
}

Now we  notice that by definition the flux $s_\alpha({\bf p})$ is
the excess of particles
which momentum changes from some value smaller than $p_\alpha$  to
another one bigger than $p_\alpha$ over the particles which momentum
changes in the opposite direction, from bigger to smaller values.

The total number of particles increasing their momentum is given by
$$
\int_{q_\alpha >0} d^3 q \int d^3 p^\prime \int_{p_\alpha-q_\alpha}^{p_\alpha}
   dp_\alpha
 w \left( {\bf p}+ {{\bf q} \over 2}, {\bf p^\prime} -{{\bf q} \over 2} ,
 {\bf q} \right) f({\bf p}) f^\prime ({\bf p^\prime})  $$

while the total number of particles decreasing their momentum is
$$
\int_{q_\alpha >0} d^3 q \int d^3 p^\prime \int_{p_\alpha-q_\alpha}^{p_\alpha}
   dp_\alpha
 w \left( {\bf p}+ {{\bf q} \over 2}, {\bf p^\prime} -{{\bf q} \over 2} ,
 -{\bf q} \right) f({\bf p}+{\bf q}) f^\prime ({\bf p^\prime})  $$

Because of the relation \Eprop \ the functions $w$ are the same in both
expressions.
We need to compute the difference of these two integrals.
According to what was said above, the most important contribution
to this expression comes from the small values of ${\bf q}$.
So we can expand everything in powers of ${\bf q}$ keeping only the
first term of the expansion while the integral over $p_\alpha$ can
be  replaced  by multiplying by $q_\alpha$.
We obtain

\eqn\Eflu {
\int_{q_\alpha >0} d^3 q \int d^3 p^\prime
w({\bf p}, {\bf p^\prime}, {\bf q}) \left( f({\bf p}) { \partial f^\prime
({\bf p^\prime}) \over \partial p^\prime_\beta }-f^\prime({\bf p^\prime})
{\partial f( {\bf p}) \over p_\beta} \right) q_\alpha q_\beta }

The integrand of \Eflu \ is an even function of ${\bf q}$ due to
\Eprop. So we can replace the integral over $q_\alpha>0$ by half the
integral over all ${\bf q}$.

Now we can introduce the collision cross section by using
\eqn\Eur { w d^3 q = d \sigma {c I \over \epsilon_1 \epsilon_2}, }
where $I$ is a well known  invariant
$$ I= \sqrt{ (p_1 p_2)^2-m_1^2 m_2^2 }, $$
$p_1$, $p_2$ are the particle four momenta, $m$ are their masses, and
$\epsilon$ are their energies. To make the calculations simpler
we are going to write all the four-vectors and to do all the relativistic
computations with the units where $c=1$ while putting $c$ explicitly
in all the kinetic formulae.
It should not result in any misunderstanding.

After substituting \Eur \ into \Eflu \ we arrive at

\eqn\Efin {s_\alpha= \int d^3p^\prime  \left( f({\bf p}) { \partial f^\prime
({\bf p^\prime}) \over \partial p^\prime_\beta }-f^\prime({\bf p^\prime})
{\partial f( {\bf p}) \over p_\beta} \right) B_{\alpha \beta} }
where
\eqn\Ebe { B_{\alpha \beta}=
{1 \over 2} \int q_\alpha q_\beta { I \over \epsilon_1 \epsilon_2} c d\sigma }

Thus far we were able to proceed without assuming anything about whether
the particles are relativistic or not and what kind of interaction
governs their collisions. Now our next task is to compute the quantities
$B$ which are of course dependent on what kind of plasma we deal with.
{}From now on, we are going to consider the case of ultrarelativistic
particles.

First, we already saw  that the distribution
$f^\prime$ describes the bulk of quarks distributed according Maxwell
distribution, while the distribution $f$ is the one of the possible
runaway quarks we will need to determine later.
 So
\eqn\Edist { f^\prime({\bf p^\prime})= C \exp(-{c p^\prime \over T}) }
where $C$ is the normalization coefficient, $c$ is the speed of light (the
quarks are ultrarelativistic!) and
$T$ is the temperature.

The tensorial structure of $B_{\alpha \beta}$ can be determined by a
very simple argument. If $f$ were also a Maxwell distribution of the
form \Edist \ it would make the flux \Efin \ to be zero.
Let us substitute \Edist \ into \Efin  \ for {\sl both} $f$ and $f^\prime$.
The integrand of \Edist \ will then become proportional to
$$ {c \over T} f f^\prime \left( {p^\prime_\beta \over p^\prime}-
{p_\beta \over p} \right) B_{\beta \alpha} $$
On the other hand, it is better be zero. From here we see that
 $B_{\alpha \beta}$ is a transverse tensor with respect to the
difference of the unit vectors lying in the direction  of the
momenta of the colliding particles,
\eqn\Ebett { B_{\alpha \beta}={1 \over 2} B \left[ \delta_{\alpha \beta}-
{\left( {p_\alpha \over p}-{ p^\prime_\alpha \over p^\prime} \right)
 \left( {p_\beta  \over p}-{ p^\prime_\beta \over p^\prime} \right)
\over
{ \left( {{\bf p}  \over p}-{ {\bf p^\prime} \over p^\prime} \right)}^2
}
\right]
}
where    $B$ is just the trace of $B_{\alpha \beta}$,
\eqn\Etr {
B={1 \over 2} \int {\bf q}^2 { I \over \epsilon_1 \epsilon_2} c d\sigma }

Another way of seeing the same is to notice that if the momentum
transfer ${\bf q}$ is a very
small vector in comparison with the momenta of colliding ultrarelativistic
particles it will be orthogonal to the difference ${{\bf p}  \over p}-{
{\bf p^\prime} \over p^\prime}$ due to the energy conservation and then
\Ebett \ follows  from \Ebe.

Now let us proceed with computing $B$ according to \Etr.
The differential cross section of the electron-electron collision
due to the photon exchange in the ultrarelativistic limit can be
taken from, for example, \RLa \  and is equal to
\eqn\Elansec { d\sigma=
{e^4 \over \epsilon^2} { {\left( 3+\cos^2{\theta}) \right)}^2 \over 4
\sin^4\theta } do
}
where $\epsilon$ is the energy of each of colliding particles in the
center of mass reference frame and $\theta$ is the standard scattering
angle in the same frame. There is an interference factor in \Elansec\
which takes into account that the electrons are identical particles, but
this factor is close to 1 at small scattering angles  $\theta
\ll 1$. The whole nature of Landau collision integral means only small
scattering angles are important so the interference terms can be neglected.

We will take the formula \Elansec\ as the quark-quark collision cross section
substituting $\alpha$ instead of $e$, $\alpha$ being the strong
interaction effective coupling constant at the energies we consider
($\alpha \approx 0.1 \dots 0.2$). So
\eqn\Ecrossdif { d\sigma=
{\alpha^2 \over \epsilon^2} { {\left( 3+\cos^2{\theta}) \right)}^2 \over 4
\sin^4\theta }do
}
By taking the cross section in the form \Ecrossdif, we assume that the
momentum of fast quarks is much larger than the Debye mass, meaning that
the typical scattering is at distances much smaller than the Debye length.

Now $do$ is the spherical angle differential in the center of mass
reference frame.  But in \Etr \ we integrate ${\bf q}^2$, which is the
transferred momentum in the laboratory frame.
In order to perform the integration, we have to express ${\bf q}^2$
in terms of the angle $\theta$.
In general, that turns out to be a long algebraic formula which
is difficult to work with. However, we can considerably
simplify the calculations by using the fact that one of the colliding
particles is a very fast quark, while the other is slow, even though
both of them is ultrarelativistic (don't forget that we consider
fast ranaway quarks in the ``sea" of the slow Maxwell ones).
In this approximation it is rather
easy to transform the ultrarelativistic momenta to the
center of mass frame and then back to express the momentum transferred
${\bf q}$ in terms of the scattering angle.

\ifig\Fglypykartinka {Quark scattering}
{\epsfxsize
5.0in\epsfbox{quark.ps}}

\Fglypykartinka \ shows the 3-momenta of fast (${\bf P}$) and slow (${\bf
p}$) quarks.  We choose the frame of reference in such a way that the
projections of ${\bf P}$ and ${\bf p}$ on the vertical axis $Y$ are the same.
Then the horizontal exis $X$ will be the axis of the boost transformation
we need
to perform to pass to the center of mass reference frame. Since
$P \gg p$, the angle between ${\bf P}$ and the $X$-axis is very small and
 the angle $\alpha$ between ${\bf p}$ and the axis $X$-axis almost
coincides with
the angle between ${\bf P}$ and ${\bf p}$. If we denote
${\bf p^\prime}$ to be
the momentum of each quark in the center of mass frame, then
from \Fglypykartinka \ it follows that the projections $p^\prime_x$ and
$p^\prime_y$ of ${\bf p^\prime}$ to the axes $X$ and $Y$ are (here we choose
the speed of light $c=1$)

\eqn\Etrans {
\eqalign {p^\prime_x &= { P - P v \over \sqrt{ 1-v^2}} \cr
         p^\prime_y &= p \sin \alpha}
}

where v is the velocity of the boost.
On the other hand,
\eqn\Etranss {
-p^\prime_x= { p \cos \alpha - p v \over \sqrt{1 - v^2}}
}
So
\eqn\Evelo {
v= {P+p \cos \alpha \over P + p} \approx 1 - {p \over P}(1-\cos \alpha)
}

Now let $q^\prime$ be the 3-momentum transfer in the center of mass frame.
Its direction with respect to ${\bf p^\prime}$
is defined  by the scattering angles. The energy transfer in this
frame is zero (the 3-momenta of the colliding particles just rotate
without changing their magnitudes).
\eqn\Egf {
q^{\prime 2}=4 p^{\prime 2}  \sin^2~{\theta \over 2}
}
$\theta$ being the scattering angle. We need to find the momentum transfer
$q$ in the laboratory frame while keeping it expressed in terms of the
center of mass scattering angles.

To do that, we compute the projections of the vector $q^\prime$ to the
axes $X$, $Y$, and $Z$ ($Z$ is perpendicular to the plane of \Fglypykartinka)
to obtain
\eqn\Egpol {
{\bf q^\prime } = p^\prime
 (\cos\theta-1) (\cos\beta, \sin\beta,0) + p^\prime \sin\theta
(-\cos\varphi~\sin\beta, \cos\varphi~\cos\beta, \sin\varphi)
}
where $\beta$ is the angle between the $X$-axis and ${\bf p^\prime}$
while $\varphi$ is the azimuthal scattering angle. One may check that
\Egpol\ square gives exactly \Egf.

The square of the four-vector $q_4$ is clearly just
$$ q_4^{2}=-{\bf q^{\prime 2}}=-4 p^{\prime 2}\sin^2 {\theta \over 2} $$
Using that we transfer ${\bf q^\prime}$ back to the laboratory frame
to obtain
\eqn\Eqlab {
\eqalign {
{\bf q}^2=-q_4^2+q_0^2&=-q_4^2+{q^{\prime 2}_x v^2 \over 1 - v^2} = \cr
4 p^{\prime 2}  \sin^2~{\theta \over 2} &+  p^{\prime 2}
{( (\cos~\theta-1) \cos~\beta - \sin~\theta~\cos~\varphi~\sin~\beta)}^2
{v^2 \over 1 - v^2} }
}

Now we use that only the small scattering angles are important,
that is $\theta \ll 1$, to get
\eqn\Eqlabone {
{\bf q}^2=  p^{\prime 2} \theta^2  \left( 1 + \cos^2\varphi~\sin^2\beta
{v^2 \over 1 - v^2} \right)
}

And the last step is to use that
\eqn\Epprimenew {
p^{\prime 2}\approx p^{\prime 2}_x=pP {1 - \cos~\alpha \over 2},
}
the formula \Evelo\
and
\eqn\Ebetata{
\sin\beta\approx {p^\prime_y \over p^\prime_x} =
\sin~\alpha \sqrt{ 2 p \over P} {1 \over \sqrt{1-\cos\alpha}}
}

Substituting \Evelo, \Epprimenew, and \Ebetata\ to the \Eqlabone\ we get
\eqn\Eqfinal {
{\bf q}^2 = \theta^2 pP { 1 - \cos\alpha \over 2}
\left( 1+\cos^2\varphi {\sin^2\alpha \over {(1 - \cos\alpha)}^2} \right)
}
This completes the task of expressing $q$ in terms of scattering
and azimuthal angles.

Before proceeding further, we can average over the azimuthal angle to get
\eqn\Eqmfinal {
{\bf q}^2 = \theta^2 pP { 1 - \cos\alpha \over 2}
\left( 1+ {\sin^2\alpha \over 2 {(1 - \cos\alpha)}^2} \right)
}

Now we can plug everything into the \Etr. Keeping in mind that
the most important angles are the small ones, $\theta \ll 1$, and using
$${I \over \epsilon_1 \epsilon_2} = 1 - \cos\alpha$$
we
arrive at
\eqn\Etrcomp { B= {4 \pi \alpha^2 \over c} (1-\cos\alpha) \left( 1+
{\sin^2\alpha \over 2 {( 1 - \cos\alpha)}^2} \right) \int {d \theta \over
\theta}
}

The integral over $\theta$ turns out to be logarithmically divergent.
The same thing happens to the nonrelativistic plasma as well
so we should not be discouraged. In fact, it should be cut off
at some small angles due to the finite density of plasma and so,
the impossibility of very small angle scattering, and it large angles
because (23) is applicable only when $\theta \ll 1$. Within the logarithmic
accuracy, it is enough to replace the divergent integral by some
number $L$ called the Coulomb logarithm. Then $B$ becomes

\eqn\Etrcomp { B= { 4 \pi \alpha^2 L \over c} (1-\cos\alpha) \left( 1+
{\sin^2\alpha \over 2 {( 1 - \cos\alpha)}^2} \right)
}

$B$ has  a simple meaning of the average momentum square transfer
in a collision. We see that $B$ does not depend on the momenta of the
colliding particles at all unlike the nonrelativistic case where $B$
decreases with the momentum as $1/P$ (see ref. \RLL). As we will see later,
that changes completely the behavior of fast quarks in comparison with fast
electrons.  While fast electrons run away in the direction of the electric
field, fast quarks do not.

Now we are in the position to derive the final expression for the
flux from \Efin. To do that, we substitute \Etrcomp\ with \Ebett\ into \Efin.
Then we remember that the function $f^\prime(p^\prime)$ is a Maxwell
distribution \Edist\ while the function $f(p)$ is the unknown distribution of
the ranaway quarks. And we integrate out the momentum $p^\prime$ of
the equation \Efin. There are two terms in \Efin\ we have to compute.
Let us begin with the first one.
$$ - \int d^3 p^\prime {\partial f({\bf p}) \over \partial p_\beta}
B_{\alpha \beta} f^\prime({\bf p^\prime}) $$
We introduce the unit vectors ${\bf n}={\bf p}/p$ and ${\bf n^\prime}=
{\bf p^\prime}/p^\prime$. In terms of them the expression becomes
$$ - 2 \pi \alpha^2 { L N \over c} {\partial f \over \partial p_\beta}
\int {do^\prime \over 4 \pi} \left( {3 \over 2} - {1 \over 2} n_\gamma
n^\prime_\gamma \right) \left[ \delta_{\alpha \beta} - {(n_\alpha -
n^\prime_\alpha) (n_\beta-n^\prime_\beta) \over 2 - 2 n_\gamma
n^\prime_\gamma} \right] $$ where the integration ${ do^\prime \over 4 \pi}$
gives the average over the spherical angle of ${\bf n^\prime}$ and we used
that the integration over the magnitute of ${\bf p^\prime}$ gives us just the
density of quark plasma:  $$ N = 4 \pi \int p^{ \prime 2} dp^\prime f^\prime
(p^\prime) $$

To compute the integral over the spherical angle we notice
that the only vector the result may depend on is ${\bf n}$.
So
$$
\int {do^\prime \over 4 \pi} \left( {3 \over 2} - {1 \over 2} n_\gamma
n^\prime_\gamma \right) \left[ \delta_{\alpha \beta} - {(n_\alpha -
n^\prime_\alpha) (n_\beta-n^\prime_\beta) \over 2 - 2 n_\gamma
n^\prime_\gamma} \right] = C_1 \delta_{\alpha \beta} + C_2 n_\alpha n_\beta$$
By taking the trace of this expression and by multiplying it by
$n_\beta$ we immediately obtain
$$
\int {do^\prime \over 4 \pi} \left( {3 \over 2} - {1 \over 2} n_\gamma
n^\prime_\gamma \right) \left[ \delta_{\alpha \beta} - {(n_\alpha -
n^\prime_\alpha) (n_\beta-n^\prime_\beta) \over 2 - 2 n_\gamma
n^\prime_\gamma} \right] = {7 \over 6} \delta_{\alpha \beta} - {1 \over 2}
 n_\alpha n_\beta$$
Therefore we are left with
$$ - 2 \pi \alpha^2 { L N \over c} {\partial f \over \partial p_\beta}
\left( {7 \over 6} \delta_{\alpha \beta} - {1 \over 2}
 n_\alpha n_\beta  \right)
$$
We will see later that we need just the radial component of that
expression. It is
\eqn\Epartflux {
- { 4 \pi \alpha^2 L N  \over 3 c }   {\partial f \over \partial p}
}
We don't have to compute the second term of \Efin\ which is proportional
to $f$. The expression for the radial component of the flux should
look like
\eqn\Enewfl {
s=C {\partial f({f p}) \over \partial p} + D f({\bf
p}) }

It turns out we do not need to compute both $C$ and $D$ in \Enewfl\
because they are related. The Maxwell distribution
must make the flux \Enewfl\ be zero. That is
$$ C {\partial \exp\left({- c p \over T}\right) \over \partial p}
+ D \exp\left( {- c p\over T} \right) =0
$$
or
\eqn\Eenewfl {
s= A \left( {\partial f(p) \over \partial p} +
{c \over T}  f(p ) \right)
}
So it is enough to compute the coefficient in front of the
term of \Efin\ with the derivative of $f$. But it had been already
computed and given by \Epartflux. Finally we obtain
\eqn\Efflll {
s=- { 4 \alpha^2 L N \over 3 c} \pi
\left( {\partial f(p) \over \partial p} +
{c \over T}  f(p ) \right)
}
The main difference between the flux of ultrarelativistic particles
we have just obtained with that of classical particles found in
\RLL\ is the momentum dependence of the ``classic'' flux.
It is $s_{\rm classic} \propto {1 \over p^3}$ while relativistic
flux does not depend on the momentum at all. That strong collision
flux will eventually prevent particles from accelerating in the direction of
the electric field.

Now, with the expression for the flux in our hands, we can proceed to
discuss the kinetic equation.

The general form of the kinetic equation is easy to write down
as soon as the expression for the flux is known
\eqn\Egenkin {
{\partial f \over \partial t} + e {\bf E} {\partial f \over
\partial {\bf p} } + {\rm div}_{\bf p} {\bf s} = 0 }
where $e$ is the charge of the particular quark and {\bf E} is  the
magnitute of the electric field.

In general the function $f$ depends on both the coordinates and the
momenta. The geometry of our problem is spherical. We have a sphere
of quark-gluon plasma surrounded by the electric charge which means
 spatial nonuniformity of the electric field. Nevertheless,
the basic question whether the quarks are going to run away or not
does not depend on any geometry. Since it is much easier to consider the
quarks in the homogeneous electric field, we are going to solve that
problem. Our spherical problem cannot give rise to a totally
different behavior, qualitatively the answer will be the same.

We conjecture now that there is such a phenomenon as runaway
quarks. That means the solution of the kinetic equation will give
us the function $f({\bf p})$ which will describe the runaway quarks
moving in the direction of the electric field. To find the flux of
such particles we can average the kinetic equation over the directions,
following \RLL.
\Egenkin\ gives us
\eqn\Emankin {
{\partial f \over \partial t} + eE {\partial \over p^2 \partial p}
p^2 f + {\partial \over p^2 \partial p} p^2 s =0}
while the  stationary solution of kinetic equation satisfies, with
the aid of \Efflll,
\eqn\Estatkin {
 eE {\partial \over p^2 \partial p}
p^2 f - {\partial \over p^2 \partial p} p^2 \left[
 { 4 \alpha^2 L N \over 3 c} \pi
\left( {\partial f(p) \over \partial p} +
{c \over T}  f(p ) \right)
\right] =0}
This is the fundamental equation we have been deriving.

Let us try to find the constant flux stationary solutions. They are given by
\eqn\Econflu {
p^2 \left[ eE f -
 { 4 \alpha^2 L N \over 3 c} \pi
\left( {\partial f(p) \over \partial p} +
{c \over T}  f(p ) \right)
\right] = n
}
where $n$ is the flux.
We can easily check that if the flux $n$ is not equal to zero, the solution of
\Econflu\ will always be {\it divergent} at $p \rightarrow 0$ making it
impossible to find such a solution which describes a constant flux at large
$p$ while smoothly connected with Maxwell distribution at small values of
$p$.  That means there are no runaway quarks, as we have promised.

We can try to analyze the zero flux solution,
\eqn\Ezeroflux {
f=e^{  - \left( {c \over T} - {3 e E c \over 4 \pi \alpha^2 L N } \right)
p }
.}
It gives us a slightly distorted Maxwell distribution.

Let us estimate the terms in \Ezeroflux\ to see how small the correction is.
$T$ is of the order 200 Mev. $E$ can be estimated
as being created by a charge about $100\times e$ at the distance of
about $\propto 10^{-13}cm$. The density $N$ is the number of quarks per unit
volume. Let us also estimate the number of quark as $10^2$ then $N \propto
10^{41} cm^{-3}$. That gives us
$$ {c \over T} \propto  10^{14} {s \over g~cm}$$
and
$$  {3 e E c \over 4 \pi \alpha^2  N } \propto  10^{13} {s \over
g~cm} $$

So it is difficult to say if
the electric field is strong enough to produce a noticible distortion
of the distribution function
on the basis of this very rough estimation.
Let us, however, notice that the whole derivation was based on the
idea that only  fast quarks are distorted while the distribution
\Ezeroflux\ would predict an overall distortion. So most probably
\Ezeroflux\ is unphysical.

In conclusion we would like to say that we have shown the phenomenon of
the runaway quarks does not exist due to the effects of the relativistic
kinematics. One can also undertake a similar calculations with $c$-quarks
which are massive enough to be almost nonrelativistic at our temperatures.
We will have the problem of nonrelativistic $c$ quarks trying to run away
in the sea of the ultrarelativistic $u$ and $d$-quarks.
This calculation turns out to be similar to  the one given in this
paper. We will not give the details of that calculation here, but we will
remark that the
result is the same, the heavy quarks do not ``run away'' as well.

\vskip 1cm

The author is grateful to I. Kogan for posing the problem and for many
discussions on its subject and also to the Physics Department of
Melbourne University where this work had been completed for their
hospitality.

\listrefs
\bye